\documentclass[option]{aastex631}
\usepackage{graphicx}
\usepackage{amsmath}
\usepackage{float}
\usepackage{natbib}

\shorttitle{Disk Material Inflates Gaia RUWE Values}
\shortauthors{Fitton et al.}

\begin{document}

\title{Disk Material Inflates Gaia RUWE Values in Single Stars}

\author{Shannon Fitton}
\affiliation{Department of Astronomy, The University of Texas at Austin, Austin, TX 78712, USA}

\author[0000-0003-2053-0749]{Benjamin M. Tofflemire}
\altaffiliation{51 Pegasi b Fellow}
\affiliation{Department of Astronomy, The University of Texas at Austin, Austin, TX 78712, USA}

\author[0000-0001-9811-568X]{Adam L. Kraus}
\affiliation{Department of Astronomy, The University of Texas at Austin, Austin, TX 78712, USA}

\section{Abstract} 
An understanding of the dynamical evolution of binary star systems, and their effects on stellar and planetary evolution, requires well-characterized binary populations across stellar ages. However, the observational resources required to find and characterize binaries are expensive. With the release of high-precision Gaia astrometry, the re-normalized unit weight error (RUWE) statistic has been shown to reveal the presence of binary systems, with RUWE values greater than 1.2 indicating the presence of a stellar companion within $\sim1\arcsec$. Our goal is to assess whether this new diagnostic, which was developed for field-age systems ($>$1 Gyr), applies to young systems; specifically, those that host circumstellar disks. With a control sample of single-star systems, compiled from high-contrast imagining surveys of the Taurus and Upper Scorpius  star-forming regions, we compare the RUWE values for systems with and without circumstellar disks. We show that the presence of a protoplanetary disk alone can result in inflated RUWE values. Based on the distribution of the RUWE for disk-bearing single stars, we suggest a more conservative single-star -- binary threshold is warranted in the presence of disk material. We place this cutoff at the distribution's 95th percentile, with RUWE $= 2.5$.\\

\section{Introduction}

Binary systems are a common product of star formation, with young populations holding a particular interest given evidence that the binary fraction is highest in the formation environment (\citealt{DucheneKraus2013}). Obtaining a better understanding of their occurrence rates and properties as a function of stellar age is critical, as binarity can affect many processes that shape the formation and evolution of both stars and planets (\citealt{hurley2002,marzari}). However, binaries can be difficult to find and characterize, as the expensive nature of conducting adaptive-optic (AO) surveys (specifically with large telescopes) is limiting. This can be largely mitigated by looking instead towards Gaia, a survey that measures the three-dimensional spatial and velocity distribution of stars (\citealt{gaiacollab}).\\

This is useful for both wide ($\rho > 1 \arcsec$) and close binaries ($\rho < 1 \arcsec$). Wide binaries are easily resolved by the Gaia point-spread-function (PSF) to 1$\arcsec$. Close binaries that are unresolved can be detected through the re-normalized unit weight error (RUWE) of the astrometric solution (\citealt{ziegler2018}). Multiple studies have now shown that RUWE values greater than 1.2 signal the presence of a stellar companion within 1$\arcsec$ and with a $G$-band contrast ratio $<$5 mags (\citealt{rizzuto2018}, Kraus et al. in prep). Additionally, this method offers the potential to identify binary populations at many distinct ages, especially as the number of nearby moving groups and associations with well-known ages also increases with the help of Gaia astrometry (e.g., \citealt{Miret-Roig,wallace,ujjwal,kounkel}).\\ 

In this Research Note we assess whether the Gaia RUWE threshold for binary detection, which was developed using field-age stars ($\tau > 1$ Gyr), applies to the youngest stellar populations. There are various possible systematic effects that are common within younger populations, such as the presence of protoplanetary disks ($\tau \lesssim 10$ Myr), and we seek to establish how these effects may influence a star's RUWE. Our analysis is motivated by the empirical sensitivity of the RUWE statistic to deviations from a ``single-star" light profile, and determining the likelihood of common binary-related issues affecting this statistic. Our primary focus is to further investigate how the presence of disk material alters this profile in a way that inflates the RUWE value, such as via scattered light that is not centered on the star. Additionally, spot checks of various disk bearing sources that are known to be single from AO monitoring have shown inflated RUWE values (e.g., AA Tau; RUWE  = 3.33 (\citealt{gaiacollab})). To systematically test this new diagnostic at young ages, we assemble a sample of single stars from AO surveys of the Taurus star-forming region ($\tau \sim 0-5$ Myr; \citealt{krolikowski2021}) and compare the RUWE distributions for sub-samples with and without circumstellar disks. From this sample we evaluate the efficacy of the field star RUWE threshold when applied to young, disk-bearing stars, and determine how best to use the RUWE statistic in the search for binary systems in young populations.\\ 

\section{Data \& Methods}
We draw our sample from studies that performed AO surveys of star-forming regions in order to look for binary companions, primarily in Taurus and Upper Scorpius. These studies report contrast limits using the F555W, F775W, and F850LP filters from the Hubble Space Telescope (\citealt{kraus2005,kraus2006}), K', $K_s$, L, and H' filters from the Keck II 10 m and Palomar Hale 200" telescopes (\citealt{kraus2008,kraus2011}), and the $K_p$, $K_s$, and $L_p$ filters from Keck NIRC2 (\citealt{cheetham2015}). From these studies we identified 154 targets that are confirmed to be binary or single. To ensure validity of our sample, we only include single stars that cite a contrast limit $>1$ magnitude at 10--40 milliarcseconds, reducing the single-star sample to 122.\\

We then compile the RUWE and WISE W1-W3 values from the Gaia EDR3 and ALLWISE (\citealt{cutri}) databases, respectively. The data are presented in Figure 1 (panel a). We set a WISE color cut of 1, where stars with W1-W3 $>$ 1 are expected to have a protoplanetary disk. Likewise, stars with W1-W3 $<$ 1 are presumed to not have disks (\citealt{Pecaut&Mamajek2013}).

\section{Results \& Conclusion}

We create a histogram of the disk-bearing and disk-free samples in order to compare the two populations (Figure 1, panel b). To test whether the two samples are drawn form the same parent sample, we conduct a KS test which results in a value of 0.385 and a corresponding p-value of 0.000276. The cumulative distribution is presented in Figure 1, panel c. This proves that young, disk-bearing single stars have higher RUWE values on average than young, disk-free stars. \\

\begin{figure*}[!ht]
    \centering
    \includegraphics[width=0.303\textwidth]{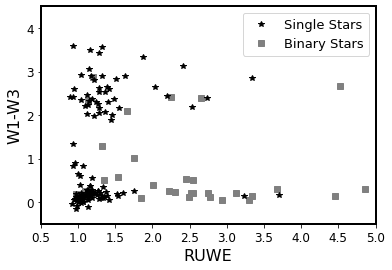}
    \includegraphics[width=0.317\textwidth]{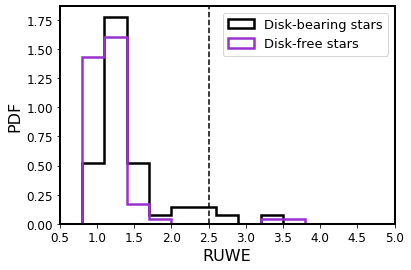}
    \includegraphics[width=0.303\textwidth]{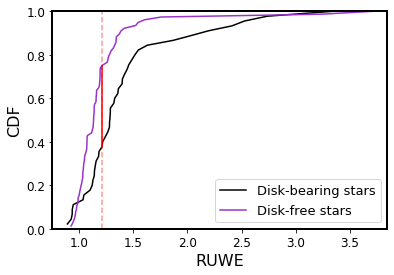}
    \label{fig:example}
    \caption{{\bf Left:} Confirmed single and binary systems in our sample. {\bf Middle:} RUWE distribution for single stars, with 95th percentile indicator for disk-bearing, single-stars sample. {\bf Right:} CDF of disk-bearing and disk-free stars, with the vertical line indicating the KS test value.}
\end{figure*}

%\begin{figure}[!ht]
%    \centering
%    \subfloat[\centering Confirmed single and binary systems in our sample. ]{{\includegraphics[width=5cm]{vettedsinglewcontrasts_scatter.png} }}%
%    \qquad
%    \subfloat[\centering RUWE distribution for single stars, with 95th percentile indicator for disk-bearing, single-stars sample.]{{\includegraphics[width=5cm]{vettedsinglewcontrasts_hist.png} }}%
%    \qquad
%    \subfloat[\centering CDF of disk-bearing and disk-free stars, with the vertical line indicating the KS test value.]{{\includegraphics[width=5cm]{kstest_withline.png} }}%
%    \label{fig:example}%
%    \caption{}%
%\end{figure}

We observe that the traditional RUWE cutoff for single stars at 1.2 is too conservative to account for single stars with disks. We calculate the 95th percentile for our disk-bearing and disk-free sample, producing values of 2.5 and 1.6, respectively. Our disk-free RUWE population is broader than the field-age sample used by \citet{bryson2020}, which has a 95th percentile of 1.15. This may be the result of contamination in our disk-free sample from either hidden companions or disk material not probed by the W1-W3 color. Based on our analysis, we suggest a RUWE cutoff value of 2.5 for the identification of binary candidates in disk-bearing sources. 

A total of 6 stars were found to have an RUWE $>$ 1.5 yet W1-W3 $<$ 1, signifying a small population of confirmed single stars with inflated RUWEs that can't be attributed to the presence of a disk or a known companion. These stars are GSC06208–00834, RXJ1602.0–2221, RXJ1606.2–2036, ScoPMS021, ScoPMS044, and USco160142.6–222923. It is worth investigating these stars further, as they hold a potential for possessing a previously undiscovered protoplanetary disk or a binary companion.\\

\section{Acknowledgements}
This work has made use of data from the European Space Agency (ESA) mission
{\it Gaia} (\url{https://www.cosmos.esa.int/gaia}), processed by the {\it Gaia}
Data Processing and Analysis Consortium (DPAC,
\url{https://www.cosmos.esa.int/web/gaia/dpac/consortium}). Funding for the DPAC
has been provided by national institutions, in particular the institutions
participating in the {\it Gaia} Multilateral Agreement.

\end{document}